\documentclass[referee]{aa}

\newcommand{\kms}{\,$\mathrm{km\, s^{-1}}$}
\newcommand{\glog}{log\,$g$\,}

\usepackage{txfonts}
\usepackage{graphicx}
\usepackage[breaklinks]{hyperref}
\usepackage[authoryear]{natbib}

\bibpunct{(}{)}{;}{a}{}{,} 

\begin{document}

\title{Li -  O anti-correlation in NGC\,6752: evidence for Li-enriched polluting gas
\thanks{Based on observations collected at the ESO  VLT. Programme --81.D-0287 }
}


\author{Z.-X. Shen\inst{1,2} \and P. Bonifacio\inst{3,4} \and L. Pasquini\inst{2}  \and
S. Zaggia\inst{5} }

\offprints{L. Pasquini, \email{lpasquin@eso.org}}

\institute{National Astronomical Observatories, Chinese Academy of
Science, 20a Datun road, Chaoyang District, Beijing, China
 \and ESO - European Southern Observatory, Karl-Schwarzschild-Strasse 2, 85748 Garching bei
M\"unchen, Germany
  \and GEPI, Observatoire de Paris, CNRS, Universit\'e Paris Diderot, Place Jules Janssen 92190 Meudon, France
  \and Istituto Nazionale di Astrofisica, Osservatorio Astronomico di Trieste, Via Tiepolo 11, 34143 Trieste, Italy
  \and Istituto Nazionale di Astrofisica, Osservatorio Astronomico di Padova, Vicolo dell'Osservatorio, Padova, Italy }

\date{Received  / Accepted }

\abstract
{Elemental correlations and anti-correlations are known to be
present in globular clusters (GCs) owing to  pollution  by CNO
cycled gas. Because of its fragility Li is destroyed at the
temperature at which the CNO cycling occurs, and this makes Li a
crucial study for the nature of the contaminating stars. }
{We observed 112 un-evolved stars at the Turnoff of  the NGC\,6752
cluster with FLAMES at the VLT to investigate the presence and the
extent of a Li-O correlation. This correlation is expected if
there is a simple pollution scenario.  }
{Li (670.8 nm)  and O triplet (771 nm) abundances are derived in
NLTE. All stars belong to a very narrow region of the
color-magnitude diagram, so they have similar stellar parameters
(T$_{eff}$, \glog).  }
{ We find that O and Li correlate, with a high statistical
significance that confirms the
  early results for this cluster. At first glance this is what is expected if a simple pollution
  of pristine gas with CNO cycled gas (O-poor , Li-poor )  occurred.  The slope of the relationship,
  however, is about 0.4, and differs from unity by over
7$\sigma$. A slope of one is the value predicted for a pure contamination model.}
{We confirm an extended Li-O correlation in non evolved stars of
NGC\,6752. At the same time the   characteristic of the
correlation shows that a simple pollution scenario is not
sufficient to explain the observations. Within this scenario the
contaminant gas must have been enriched in Li. This would rule out
massive stars as main polluters, and favor the hypothesis that the
polluting gas was enriched by  intermediate or high-mass  AGB
stars, unless the former can be shown to be able to produce Li.
According to our observations, the fraction of polluting gas
contained in the stars observed is a considerable fraction of the
stellar mass of the cluster.  }


\titlerunning{Li and O  in NGC\,6752}
\authorrunning{Z. Shen  et al.}
\maketitle

\section{Introduction}
\label{sec:Intro}

Among the light elements Li is special, because Li is  produced in
the Big Bang nucleosynthesis, enriched during the Galaxy evolution
and destroyed in the stellar interior.

After the pioneering works with 4m class telescopes (Pasquini and
Molaro 1996, 1997,  Castilho et al. 2000), the study of Li in
globular clusters (GC) has received a tremendous boost with 8m
telescopes (\citealt{Boesg98}, \citealt{thevenin}, Bonifacio et
al. 2002, 2007,  Pasquini et al. 2004), and,
 with the advent of multi object spectrographs,
many un-evolved and evolved stars have been observed for Li in the nearest
GCs \citep{korn,lind,GH09,dorazim4,dorazi47tuc,monaco}.

This work shall concentrate on the GC chemical inhomogeneities.
Chemical inhomogeneities in globular clusters have been  known for
almost 50  years (Harding 1962) and Li  may represent a key
element in distinguishing between the possible mechanisms causing
them.

Historically, the first NTT observations of two Turn-Off (TO)
stars in the metal rich, chemically inhomogeneous GC 47\,Tuc did
not show any difference in their Li content, even though one star
was CN rich and the other CN poor (Pasquini and Molaro 1997). An
early claim made by Boesgaard et al. (1998) of Li inhomogeneity in
M92 has been questioned on the basis of a more realistic treatment
of the errors (Bonifacio 2002). This situation has dramatically
changed after the analysis of nine TO stars in NGC\,6752, because
for the first time strong Li variations were detected among these
stars, and in addition they were anti-correlating with Na
abundances; for seven stars an O measurement was also possible,
showing a possible O-Li correlation (Pasquini et al. 2005) .

It has been firmly established that chemical inhomogeneities in GC
bring the signatures of CNO cycling, but all the possible
reactions that involve the CNO cycle occur at temperatures well
above 10 million degrees. At these temperatures, proton captures
have destroyed all Li; the reaction destroying Li is indeed
already very effective at 2.5 MK and has an extremely high power
with temperature.  The discovery of Li variations in NGC\,6752 and
their possible anticorrelation with processed material, but with
the important detail that Li is still detected in the stars with
the lowest O abundances, therefore suggests that the stars have
been enriched by processed gas, possibly produced by the AGB stars
in their H-burning shell phase. At these temperatures also the
full CNO cycling  takes place, and oxygen is expected to be burnt.
Indeed, an oxygen-sodium anticorrelation is observed in this
cluster \citep{gratton,carretta}. We note that the NGC\,6752 TO
stars have metallicity ([Fe/H]$\sim$-1.4) and effective
temperatures (Teff$\sim$6300 K) hot enough that they should be
considered bona fide Spite plateau stars\citep{spite}, for which
internal Li depletion is excluded on the basis of standard models.
If we were able to establish on a quantitative basis the existence
of the expected Li-O correlation, we could quantitatively test the
predictions of the main existing scenarios: these stars are the
components of a second generation of low-mass stars in the
cluster, which formed from the ejecta either of intermediate mass
stars expelled during the AGB phase, as proposed by D'Antona et
al. (2002), or by material processed by massive stars  (Decressin
et al. 2007).

We are, on the other hand, conscious that not all clusters may
have gone through the same formation process. NGC\,6397 is showing
so far a puzzling behavior: in this cluster Li is almost constant
in the  observed TO stars and at the Spite plateau level, except
for a mild decrease with decreasing effective temperature
\citep{GH09}, while some stars show up to 0.6 dex  difference in
their O abundance (Pasquini et al. 2004). O-poor-Li-rich stars
have a puzzle in the pollution scenario, and only the production
during the AGB phase would at least qualitatively explain it.

In a simple self-pollution scheme Li is a powerful discriminator,
and it becomes even more powerful when used in conjunction with O,
(rather than Na or N) because both elements are destroyed in the
full CNO cycle. {\it For a given fraction of destroyed O at least
the same fraction of Li must have been destroyed.}

\section{Observations and analysis}
\label{sec:Obs}

The observations were obtained with FLAMES and the GIRAFFE
spectrograph located at the VLT Kueyen (Pasquini et al. 2003). The
Li observations were obtained in ESO Period 73 and the O IR
triplet observations in ESO period 75 for the same stars. The
resolving power and S/N  ratio obtained were of  $R\sim17,000$,
$S/N \sim 80-100$ for the Li (607.8 nm) region, and for  O I
(777.1-777.5 nm), $R\sim18,400$ and $S/N \sim 40-50$.

Targets were selected on the basis of their magnitudes and color
and on their distance from the cluster center. The BV photometry
is based on the data collected in the framework of the EIS
Pre-Flames survey \citep{Momany01}. The magnitudes were extracted
with DAOPHOT and calibrated against the Stetson photometric
sequence present in the same field. The following selection
criteria were used photometrically: stars must be in the TO region
with a magnitude range $17.0<V<17.3$ and color range
$0.42<(B-V)<0.46$; spatially,  in a radius $4<r<9$ arcmin from the
center of the cluster in order to limit the effect of crowding and
background light contamination. Furthermore, each selected star
has been visually inspected to guarantee the absence of nearby
bright objects within 5 arcsec. With this restricted limit in
magnitudes and colors the physical parameters (T$_{eff}$, \glog)
of the targets are very similar. This shall ensure that the large
chemical abundance differences observed cannot be attributed to
different stellar parameters. The color-magnitude diagram of
NGC\,6752 highlighting the selected stars is shown in Fig. 1. The
standard deviation  in $(B-V)$ color of our sample is 0.008
magnitudes, consistent with our error estimate of 0.01 mag. We
adopted a reddening $E(B-V)=0.04$\,mag \citep{harris} and computed
effective temperatures from the IRFM based calibration of
\citet{JB}. The mean effective temperature of our sample is
6353\,K with a standard deviation of 38\,K. As discussed above,
this is entirely accounted for by the photometric error. We thus
assumed all stars to have this temperature. For the surface
gravity we assumed \glog = 4.0.

Li and O triplet equivalent widths were then measured by fitting
Gaussian profiles. The Li abundances were computed with the
fitting function of \citet{sbordone10}, which is a result of the
NLTE synthesis and 3D hydrodynamical models of the CIFIST grid
\citet{ludwig}. Unfortunately a similar fitting function for the O
triplet is not yet available. We therefore computed the NLTE O
abundances with the Kiel code \citep{steen84}, a model atmosphere,
computed with the Linux version \citep{Sbord2004} of the ATLAS 9
code\citep{Kurucz93,Kurucz05}, no overshooting and the 1\,\kms\
opacity distributions functions of \citet{CK}, and the model atom
described in \citet{paunzen}. The collisions with neutral H are
treated in an approximate way (see \citealt{steen84}), and we
adopted  a scaling factor of 1/3 as recommended by \citet{caffau}.
The only \ion{O}{i} line that could be safely measured was the
777.1\,nm line, the strongest of the triplet. If the other lines
of the triplet could also be measured they implied oxygen
abundances that could disagree by as much as 0.5\,dex from the
strongest line. We therefore believe it is safer and more
homogeneous to derive oxygen abundances only from this line. We
adopt the  solar abundances of \citet{Caffau2010SoPh}.

It has long been known that the behavior of the \ion{O}{i}
permitted triplet lines is a challenge to our comprehension. For
example, Spite \& Spite (1991) compared the oxygen abundances
derived from the triplet and from the forbidden [\ion{O}{i}] lines
in LTE, finding discrepancies between 0.35\,dex and 0.75\,dex. We
cannot use the [\ion{O}{i}] lines because they are not covered by
our spectra. And the lines should have equivalent widths (EWs) of
less than 0.1\,pm, which is beyond the measurement capabilities
for stars of this magnitude with VLT. However, Garc{\'{\i}}a
P{\'e}rez et al. (2006) have studied in detail a sample of bright
subgiant stars and compared the oxygen abundances derived from
different indicators, including the \ion{O}{i} triplet, computed
in NLTE, and the [\ion{O}{i}]. They found on average an offset of
0.19\,dex, the triplet yielding a higher abundance. It is likely
that other effects, such as granulation, are the cause of this
discrepancy, and work is in progress to address this with CO5BOLD
hydrodynamical models (Ludwig et al. 2009). For the purpose of
this letter we stress that only the relative oxygen abundances are
relevant. The narrow range in effective temperature and surface
gravity covered by our sample strongly argues in favor of the
notion that the observed variations in EWs of the O I lines are
caused by oxygen abundance rather than to other effects.

The errors on both Li and O are dominated by errors in effective
temperature and in equivalent width measurement, because the
errors due to surface gravity and microturbulence are negligible
with respect to the former. We adopt an error of 100\,K on the
effective temperatures, which corresponds roughly to the
semi-dispersion in the effective temperatures determined for our
sample. The errors on equivalent widths were estimated with
Cayrel's formula \citep{Cayrel88}. The total error on abundance is
obtained by summing under quadrature the errors associated to
effective temperature and equivalent width. These amount to
0.14\,dex for oxygen and 0.09\, dex for lithium (with opposite
sign for the two elements).

Our results are displayed in Fig.\,\ref{fig:liox}; the stars for
which both oxygen and lithium could be measured are shown as
filled symbols. Open symbols with arrows are for the stars for
which one or both measures are upper limits. We have 77
measurements of both elements, for 25 stars lithium is measured,
but oxygen is not, for 4 stars oxygen is measured and lithium is
not, finally for 6 stars neither element could be measured. This
makes a total of 112 measurements and upper limits.

The visual  impression is that of a real correlation between the
abundances of oxygen and lithium, as found by \citet{pas2005},
albeit in presence of a large scatter. This is indeed confirmed by
the statistical analysis of the data. If we confine the sample to
the 77 stars for which both elements are measured, the
non-parametric test of Kendall's $\tau$ tells us that there is a
98\% probability that the two are positively correlated. To see if
the upper limits may also bring some information we used survival
statistics, as provided by
 {\tt asurv Rev 1.2}
 \footnote{\url{http://astrostatistics.psu.edu/statcodes/asurv}} (\citet{baas}).
The generalized version of Kendall's $\tau$ \citep{brown}, as
described by \citet{isobe} provides a probability of correlation
of 99.7\%, reflecting the fact that the upper limits are mostly
found along the correlation at low values of A(Li) and [O/Fe].

After establishing that there exists a correlation between [O/Fe]
and A(Li), one may venture to perform parametric fits. A
least-square fit using errors in both variables provides a slope
of 0.40 with an error of 0.08 and an intercept of 2.37\,dex with
an error of 0.01\,dex. This slope corresponds roughly to what
would have been guessed by eye and is shown as a dotted line in
Fig.\,\ref{fig:liox}. What is interesting is that the root mean
square deviation about this relation is 0.12\,dex, which is even
slightly lower than the sum under quadrature of the errors on
oxygen and lithium. This strongly suggests that in reality there
is a very tight relation between [O/Fe] and A(Li) and that the
observed scatter is only due to observational errors.

The slope of the relationship is important because, as mentioned
in the introduction, the simple pollution scheme by CNO processed
gas implies that Li and O shall diminish at least at the same
rate. Therefore the slope should be one. This is clearly not the
case, and the formal slope found by our fit is smaller, in fact a
slope of unity is at over 7 $\sigma$ from our best-fit value,
therefore the case for a slope of one can be ruled out with very
high confidence. We stress that this result does not depend on the
unique effective temperature adopted for all stars; the same
result is found when adopting a different temperature scale (based
on
 H$\alpha$ wings, Shen et al. 2010, in preparation).

\begin{figure}
\centering
\includegraphics[width=9cm,clip=true]{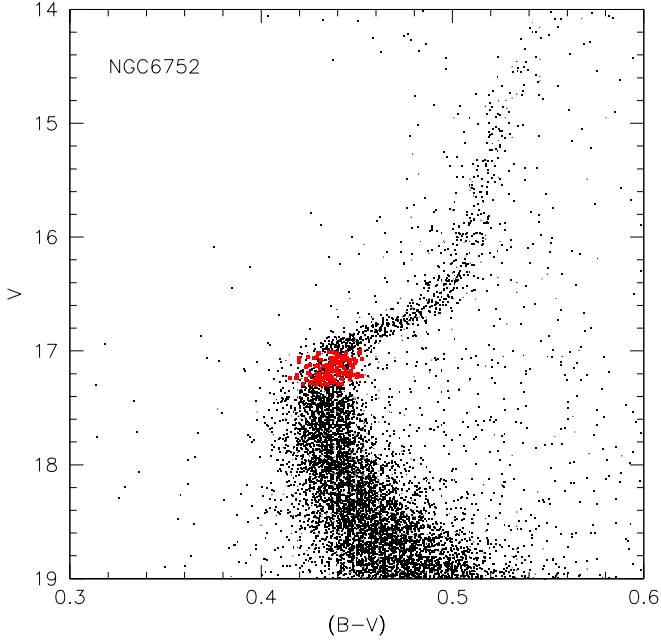}
\caption{Color-magnitude diagram of NGC\,6752. The target stars
are highlighted. Note the restricted range in magnitudes.}
\label{fig:cm}
\end{figure}

\begin{figure}
\centering
\includegraphics[width=\hsize,clip=true]{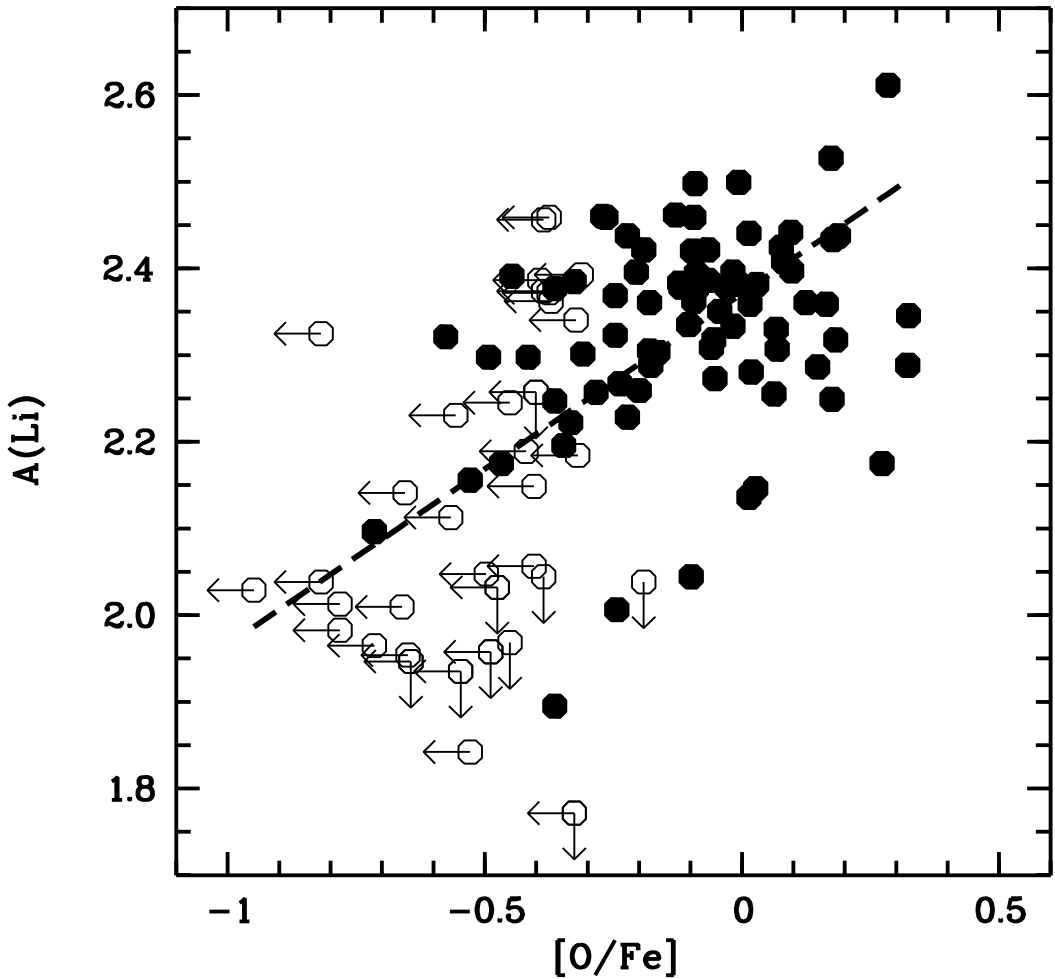}
\caption{ Li  vs. O abundance for the observed stars.}
\label{fig:liox}
\end{figure}

\section{Discussion}

We can confirm in this study a Li-O correlation for NGC\,6752.
Such a neat  correlation (together with the Li-Na anticorrelation)
is not observed in NGC\,6397 ([Fe/H]$= -$2.03; \citet{gratton})
and 47Tuc ([Fe/H]$= -$0.64; \citet{carretta04}), the other two
clusters observed in a similar way. The correlation of NGC\,6752
is close to what one would expect if a pollution scheme was at
work. If O decreases by 0.8 dex without Li - enriched polluting
material,  then Li must decrease by at least the same quantity.
Because we observe less Li depletion than O depletion, some Li
must be produced by the contaminating stars, i.e. the polluting
material {\em cannot be Li-free}.

Another evidence for this conclusion is given by the observations
of O-poor - Li-rich stars in NGC\,6397 \citep{pas2004}. The most
recent FLAMES results \citep{dorazi47tuc} show very large
variations of Na and O for stars with the same Li abundances,
pointing again to a polluting material with Li content. Our sample
also shows in Fig.\,\ref{fig:liox} a fairly large spread of O
abundance for the same Li and suggests the presence of O-poor -
Li-rich stars,  but only for one star the deviation from the
linear relationship is statistically significant. If the presence
of O-poor - Li-rich stars were to be confirmed by higher S/N ratio
observations, this would provide strong constraints on the maximum
Li abundance in the ejecta. It is interesting to note that the Li
enrichment is in general not so extreme as to destroy the general
relationship, but the  polluting gas should not have been
homogeneous and well mixed in the whole cluster. Not all stars are
in fact affected in the same way:  only for a considerable
fraction of them (5-10$\%$ in our sample), on the other hand, the
gas polluting these stars should have been very Li-rich. If we
consider a star depleted by 0.7 dex in O, this implies a
percentage of polluted gas of 80$\%$ and only 20$\%$ of pristine
one.

The star with the highest lithium abundance has a value that is
consistent with the prediction of the standard Big Bang
nucleosynthesis and the measurement of the baryonic density of the
WMAP satellite \citep{cyburt}. We believe this to be fortuitous,
because we are invoking Li-polluted material to have contributed
to the atmospheres of the studied stars and we are also stating
that nothing can be firmly said on the ``pristine'' abundances  of
the elements affected by pollution. Although in general we may
expect the Li content to be {\em lower} than the original value,
there is no strong reason to say that, occasionally, it may not be
{\em higher}, at least until the properties of the polluting stars
are firmly established.

One problem pointed out in several occasions for the  formation
from AGB ejecta is that the amount of 'polluted' material required
is huge, and from our sample we can confirm that from the oxygen
data of NGC\,6752 the amount of polluted gas is comparable to half
of the gas that formed the observed stars  (Shen et al. 2010, in
preparation). This can be seen in Figure 2: in this scheme,  all
 stars with [O/Fe]$< \sim $0.0 are composed by at least half of
the ejected material. This would pose serious constraints on the
cluster IMF. Our sample should be unbiased as far as pollution is
concerned, so it should represent the average chemical
distribution of the cluster.

We believe that any single pollution-dilution scheme we assumed
could be too simplistic,  and that the formation scenario was
likely more complex.

\begin{acknowledgements}
Z.X.S. acknowledges Chinese NSFC Grant 10903012.
\end{acknowledgements}

\bibliographystyle{aa}
{}

\end{document}